\documentclass[conference]{IEEEtran}
\IEEEoverridecommandlockouts
\usepackage{cite}
\usepackage{amsmath,amssymb,amsfonts}
\DeclareMathOperator*{\argmax}{arg\,max}

\newcommand{\norm}[1]{\left\lVert#1\right\rVert}

\usepackage[ruled,vlined]{algorithm2e}
\SetKwRepeat{Do}{do}{while}
\SetKwInput{KwInput}{Input}
\usepackage{footnote}
\usepackage{caption}
\usepackage{multirow}
\usepackage{multicol}
\usepackage{graphicx}
\usepackage{subfig}
\usepackage{textcomp}
\usepackage{xcolor}
\def\BibTeX{{\rm B\kern-.05em{\sc i\kern-.025em b}\kern-.08em
    T\kern-.1667em\lower.7ex\hbox{E}\kern-.125emX}}
\renewcommand{\baselinestretch}{1.5}
\begin{document}

\title{Water Level Estimation Using Sentinel-1\\Synthetic Aperture Radar Imagery And Digital Elevation Models
}

\author{
\IEEEauthorblockN{1\textsuperscript{st} Thai-Bao Duong-Nguyen}
\IEEEauthorblockA{\textit{University of Science} \\
\textit{Vietnam National University Ho Chi Minh City}\\
Ho Chi Minh City, Vietnam \\
1612840@student.hcmus.edu.vn}\\
\IEEEauthorblockN{3\textsuperscript{rd} Phong Vo}
\IEEEauthorblockA{\textit{University of Science} \\
\textit{Vietnam National University Ho Chi Minh City}\\
Ho Chi Minh City, Vietnam \\
phong.vodinh@gmail.com}\\
\and
\IEEEauthorblockN{2\textsuperscript{nd} Thien-Nu Hoang}
\IEEEauthorblockA{\textit{University of Science} \\
\textit{Vietnam National University Ho Chi Minh City}\\
Ho Chi Minh City, Vietnam \\
1612880@student.hcmus.edu.vn}\\
\IEEEauthorblockN{4\textsuperscript{th} Hoai-Bac Le}
\IEEEauthorblockA{\textit{University of Science} \\
\textit{Vietnam National University Ho Chi Minh City}\\
Ho Chi Minh City, Vietnam \\
lhbac@fits.hcmus.edu.vn}
}

\maketitle

\begin{abstract}
Hydropower dams and reservoirs have been identified as the main factors redefining natural hydrological cycles. Therefore, monitoring water status in reservoirs plays a crucial role in planning and managing water resources, as well as forecasting drought and flood. This task has been traditionally done by installing sensor stations on the ground nearby water bodies, which has multiple disadvantages in maintenance cost, accessibility, and global coverage. And to cope with these problems, Remote Sensing, which is known as the science of obtaining information about objects or areas without making contact with them, has been actively studied for many applications. In this paper, we propose a novel water level extracting approach, which employs Sentinel-1 Synthetic Aperture Radar imagery and Digital Elevation Model data sets. Experiments show that the algorithm achieved a low average error of 0.93 meters over three reservoirs globally, proving its potential to be widely applied and furthermore studied.
\end{abstract}


\section{Introduction}
It is undeniable that hydrological cycles have been redefined by the presence of hydropower dams and reservoirs. While reservoirs store freshwater and make it available to domestic, industrial, electricity and irrigation, dams control and manage the inflow and outflow that are reservoirs level parameters. For example, during a flood, the opening of dams should be sufficient to ensure that the capacity of the reservoir does not exceed the limits that can cause severe effects to the lower region. Monitoring the information of water in reservoirs helps in planning and managing water resources as well as drought and flood forecasting by anomaly detection.

The water level in reservoirs is traditionally measured by in-situ gauge stations installed nearby river, bridge, weir. The problem of the method is that it is immensely scarce due to its difficulties in the settings. On the other hand, even in places where gauge station exits, measured data is not always freely accessible due to national privacy policies in many countries. The alternative which can solve said limitations is remote sensing imagery.

Remote sensing has been widely used in many applications for agricultural, hydrological, disaster forecasting, etc. The coverage of many types of satellite imagery is global, which can deal with the drawbacks of in-situ gauge stations. Thanks to the European Space Agency (ESA) and Google Earth Engine \cite{Gorelick2017GoogleEveryone}, we are able to utilize the high-resolution Sentinel-1 dataset altogether with other auxiliary datasets to monitor and predict reservoirs' water levels in Greater Mekong Subregion, with the help of machine learning.

The work is organized as follows. In section \ref{relatedworks} we describe the current state of water level estimating algorithms with different kinds of sensors. Section \ref{method} explains our new idea to cope with the limitations of remotely sensed images in order to produce a robust water level estimation. Next, section \ref{experiment} shows the experiments used to assess the performance of proposed procedure. Finally, in section \ref{conclusions} we highlight the results obtained and suggest new directions for improvements.
\section{Related Works}
\label{relatedworks}
Due to the inherent ability to measure surface elevation, satellite altimetry has been extensively used to estimate water level at large reservoirs \cite{Alsdorf2007MeasuringSpace}. Because of that, numerous studies have been conducted on the use of altimetry data for water level monitoring (\cite{Frappart2006WaterData} \cite{Cretaux2011SOLS:Data}). However, altimetry data lacks global coverage as they only measure elevations along the satellite orbit tracks. Additionally, altimetry products become unreliable for small to medium sized reservoirs (less than 100 $km^2$), which the majority of dams in the Greater Mekong Subregion are.

Many studies have dedicated to mapping the water surface extent taken from optical imagery to water level, using an accurate reference DEM. Tseng et al. \cite{Tseng2016IntegratingDam} used the MNWDI band from Landsat TM/ETM+ images to extract reservoir water body then employ an Generalized Extreme Value Fitting Function to estimate the water elevation based on topographical elevations along said water body. Likewise, R{\'{e}}mi et al. \cite{RemiFLOODDATA} extracts water extent based on Landsat NVDI band, then take the average of surface heights along the water extent as the estimated water level. As optical images are sensitive to cloudy scenes, these methods tend to be insufficient for numerous real-life situations where the frequent and reliable estimations are demanded.

\begin{figure}
    \centering
    \subfloat[2014-12-15, 185.66m]{\includegraphics[width=4cm]{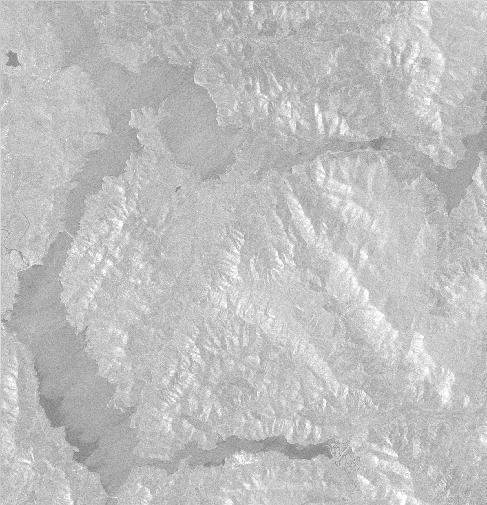}}
    \hfill
    \subfloat[2016-07-29, 186.047m]{\includegraphics[width=4cm]{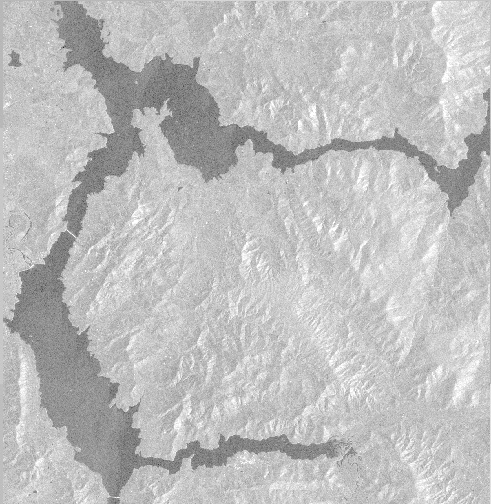}}
    \caption{Band VV of Hume dam (Australia) on two different dates with the equivalent water levels but dissimilar water surface shades.}
    \label{fig:hume_2dates}
\end{figure}

\begin{figure}
    \centering
    \subfloat[Hume dam (Australia), 2015-06-29]{\includegraphics[height=5cm]{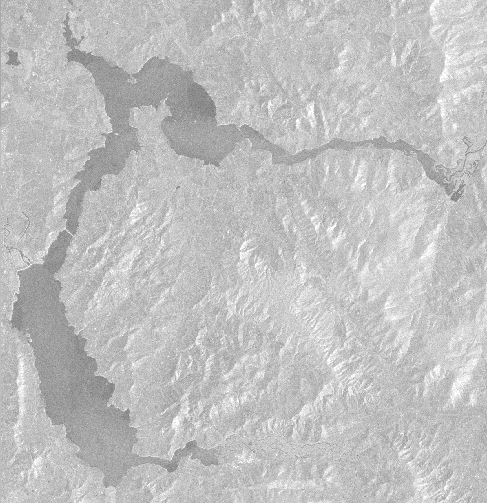}}
    \hfill
    \subfloat[Thartar dam (Iraq),
    2015-06-25]{\includegraphics[height=5cm]{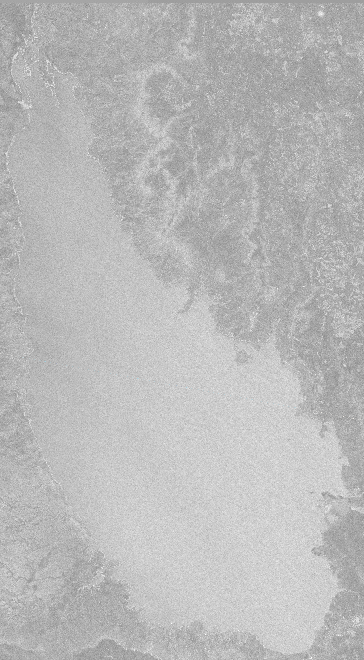}}
    \caption[Band VV of two distant dams within the same period show different gray level variations]{Two distant dams within the same period show different gray level variations. Most noticeably, Thartar dam water surface even has higher gray level than that of its land.}
    \label{fig:hume_thartar}
\end{figure}

\begin{figure}
    \subfloat[]{
        \includegraphics[width=0.3\linewidth]{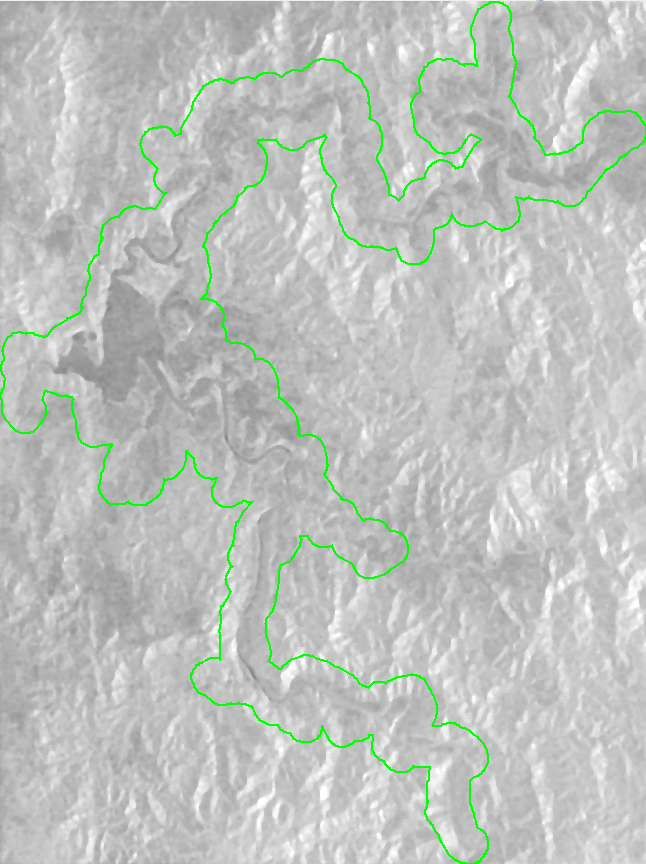}
    }
    \hfil
    \subfloat[]{
        \includegraphics[width=0.3\linewidth]{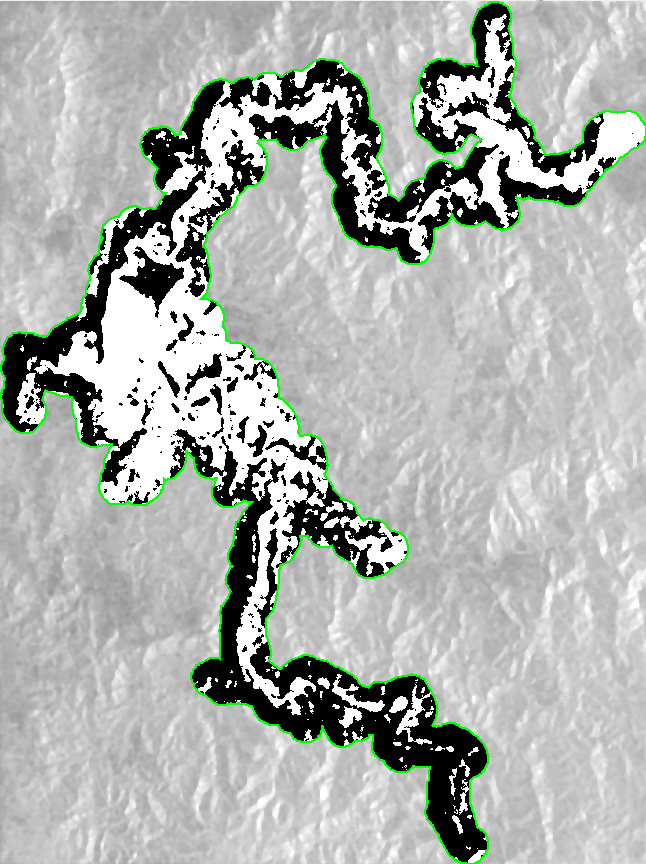}
    }
    \hfil
    \subfloat[]{
        \includegraphics[width=0.3\linewidth]{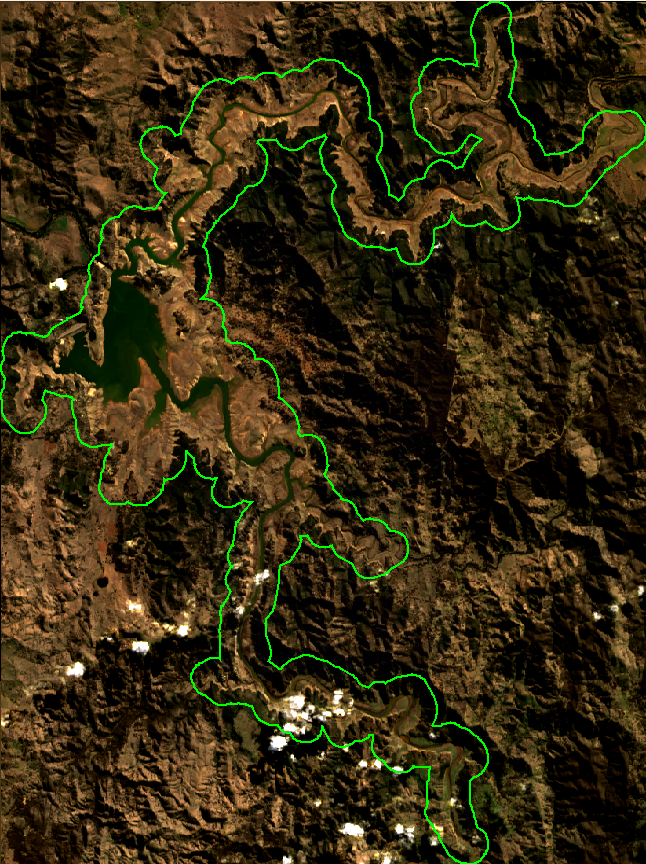}
    }
    
    \subfloat[]{
        \centering
        \includegraphics[width=\linewidth]{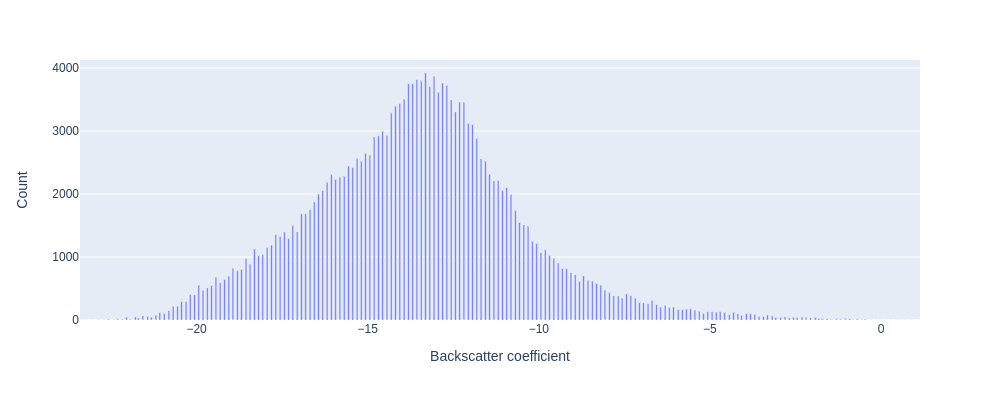}
    }
	\caption[Dam Burrendong on date 2019-07-14 when the water surface and land are difficult to be separated due to low inter-class variance.]{Dam Burrendong on date 2019-07-14 when the water surface and land are difficult to be separated due to low inter-class variance. a) Sentinel-1 image, band VV (green line: dam's shapefile). b) Otsu method's classification (white: water, black: land). c) Latest Landsat 8 image, true color visualization. d) Backscatter coefficients histogram.}
	\label{fig:badcase}
\end{figure}

Recently, some works have innovated the use of Synthetic Aperture Radar (SAR) imagery for their cloud penetration capability \cite{Hostache2009WaterFloods}. They heavily rely on the water body extraction process as the prerequisite of estimating water level. There are two main types of method employed to achieve this, both are based on thresholding the SAR image. Thresholding is based on the observation that water surfaces typically have lower back scatter coefficients than non-water areas, meaning they appear in a darker shade. For this reason, the first group of methods involve a constant greyscale threshold on an SAR image which is used to separate land and water regions. Tuan Anh et al. \cite{TuanAnh2019PREDICTINGAPPROACH} chose -17 and -22 as the threshold for VV and VH bands, respectively. Park et al. \cite{Park2020ASAR} also use SAR images and DEM to derive water level, but their water masking scheme uses a manually-inspected constant threshold; it obtains RMSE errors ranging from 1.16 to 5.25 meters over 6 different sites. We found that the dynamical range of the back scatter coefficients varies with regard to time (figure \ref{fig:hume_2dates}) and location on earth (figure \ref{fig:hume_thartar}). For the same reason, researchers have attempted to automatically choose a threshold for each independent scene. The most popular automatic thresholding approach is Otsu's method \cite{Otsu1979THRESHOLDHISTOGRAMS.}, which investigates the histogram of gray level and finds the split that maximizes the between-class variance. However it is worth noticing that thresholding methods, in general, are based on the assumption that water and land back scatter coefficient variations are significantly low, while between-class variance is high. We argue that the assumption is weak for numerous cases and hence they can not be largely employed. Figure \ref{fig:badcase} shows that in some cases, the water and/or land areas are heterogeneous in SAR back scatter coefficients.

\section{Method}
\label{method}
Here we present a novel approach for the water level estimation problem. Since water extent extraction is unreliable, we do not follow this direction. Our algorithm is based on the observation that despite the dynamic range of back scatter coefficients, the change, or the gradient magnitude, of the SAR image along the true water boundary, is the largest.
For that reason, our algorithm simulates the water rising process on the DEM image and choose the level that meets the condition above (equation \ref{eq:opt_problem}).

\small
\begin{equation}\label{eq:opt_problem}
    \begin{split}
        level^* &:= \argmax_{min_{DEM} \leq level \leq max_{DEM}}Fitness(level) \\
        Fitness(level) &:= \sum_{p \in Shoreline(level)}\norm{\nabla_{p}SAR}
    \end{split}
\end{equation}
\normalsize

Additionally, most of the computations are performed within a preliminary boundary called shapefile, provided by the Global Reservoir and Dam Database (GRanD) \cite{Lehner2011High-resolutionManagement}. Furthermore, to tackle the cases when the flood grows bigger than the shapefile, it is expanded by 500 meters further in every direction.

We implement the whole algorithm on the Google Earth Engine platform \cite{Gorelick2017GoogleEveryone}, where many of the essential remote sensing operations and data set are available for free use. Most importantly, we rely on the ability of parallel computation of Google Earth Engine to perform the water level sampling step (explained later).

First of all, there are 3 essential steps involved in the pre-processing phase.

\begin{enumerate}
    \item Firstly we combine 2 bands VV and VH of a SAR image into one band by multiplying them altogether. This is because each band may have different qualities for a given scene, sometimes VV is clearer than VH and sometimes vice versa. By combining them, we enable the clearness of both. Figure \ref{fig:NamNgumVVVH} and \ref{fig:BurrendongVVVH} show two cases when one band is better than another. Figure \ref{fig:hume_preprocess} demonstrates the intermediate resulting images of each pre-processing step for dam Hume (147.05 lon, -36.09 lat, Australia) on date 2018-11-16.
    \begin{figure}
        \centering
        \subfloat[VV band]{\includegraphics[width=4cm]{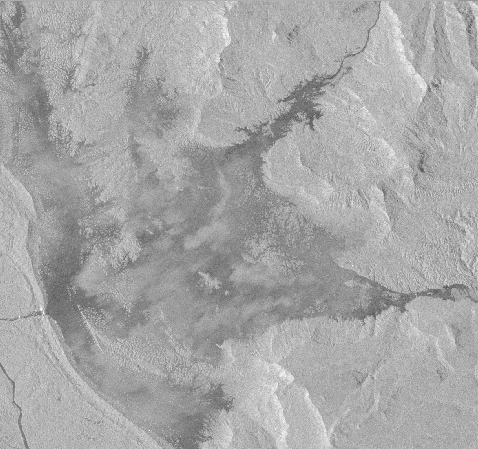}}
        \hfill
        \subfloat[VH band]{\includegraphics[width=4cm]{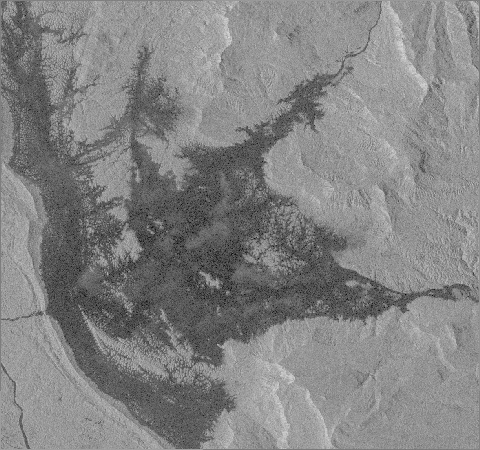}}
        \vspace{0pt}
        \subfloat[Combined band]{\includegraphics[width=4cm]{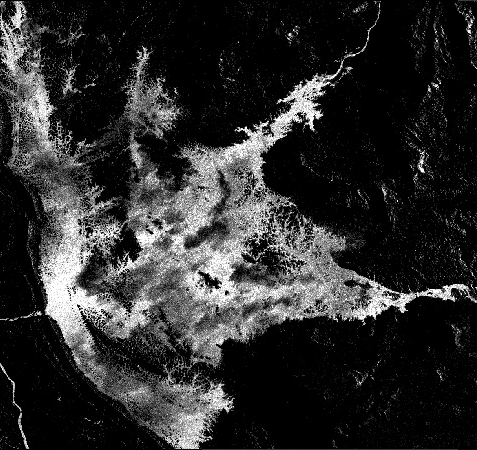}}
        \hfill
        \subfloat[Landsat 8 image]{\includegraphics[width=4cm]{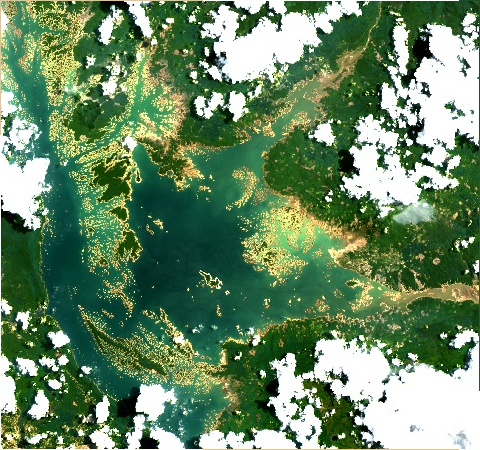}}
        \caption[Earth observation of dam Nam Ngum (102.66 lon, 18.58 lat) on date 2019-07-24 over band VV, VH and combined band from Sentinel 1 SAR image and Landsat 8 true color visualization]{Earth observation of dam Nam Ngum (102.66 lon, 18.58 lat) on date 2019-07-24 over band VV, VH and combined band from Sentinel 1 SAR image and Landsat 8 true color visualization. Band VV suffers from unknown noise that disables us to separate land and water, while band VH is much more separable. The combined band, despite inherits noises from band VV, still retains the separability from band VH. Latest Landsat 8 image to that date is shown as reference.}
        \label{fig:NamNgumVVVH}
    \end{figure}
    
    \begin{figure}
        \centering
        \subfloat[VV band]{\includegraphics[width=4cm]{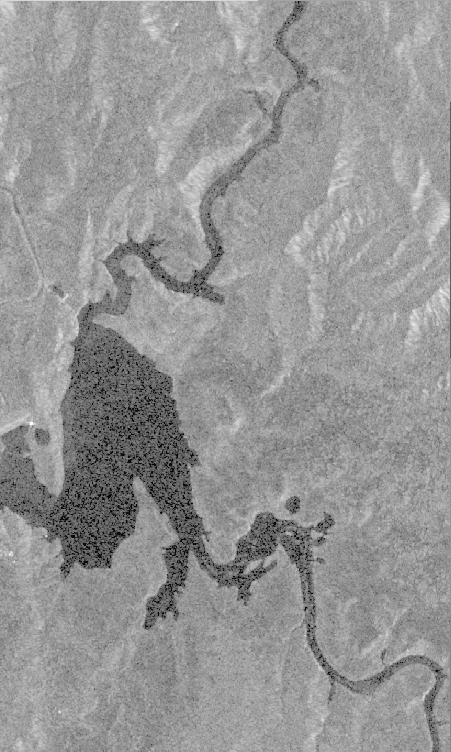}}
        \hfill
        \subfloat[VH band]{\includegraphics[width=4cm]{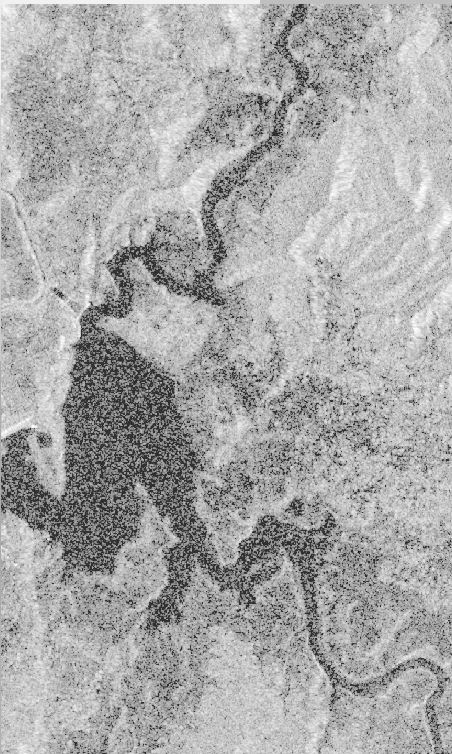}}
        \vspace{0pt}
        \subfloat[Combined band]{\includegraphics[width=4cm]{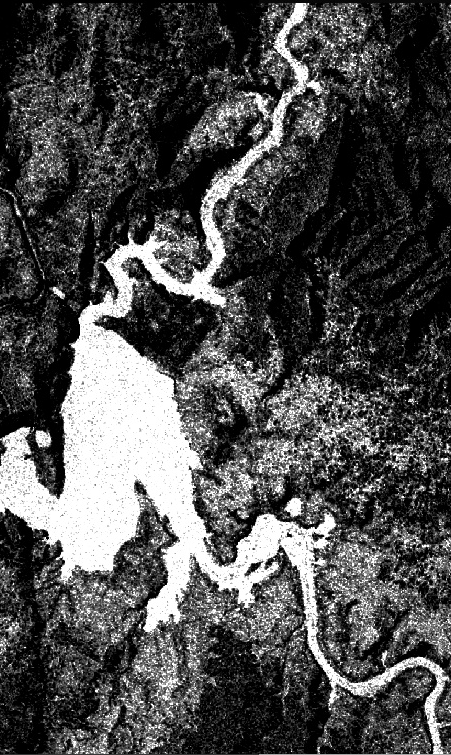}}
        \hfill
        \subfloat[Landsat 8 image]{\includegraphics[width=4cm]{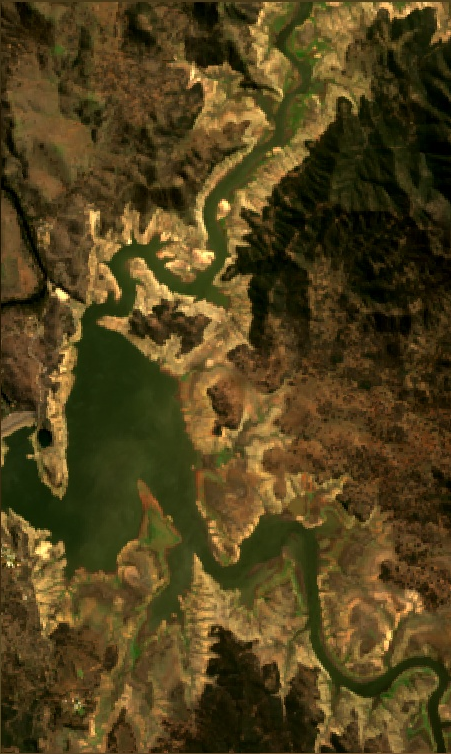}}
        
        \caption[Earth observation of dam Burrendong (149.13 lon, -32.67 lat) on date 2019-03-28 over band VV, VH and combined band from Sentinel 1 SAR image and Landsat 8 true color visualization]{Earth observation of dam Burrendong (149.13 lon, -32.67 lat) on date 2019-03-28 over band VV, VH and combined band from Sentinel 1 SAR image and Landsat 8 true color visualization. Band VH suffers from great amount of salt noise, while band VV is significantly clearer. The combined band in this case contains almost no salt noise from VV band. Latest Landsat 8 image to that date is shown as reference.}
        \label{fig:BurrendongVVVH}
    \end{figure}

    \item Secondly, we perform speckle filtering on the combined image, since SAR images have been known for being greatly subjected to salty noise. We use a focal median filter with a circle kernel for this purpose. The choice of kernel size is calibrated later.
    
    \item Thirdly, we compute the edges image using the Canny Edge Detection algorithm \cite{Canny1986ADetection} with gradient magnitude threshold of half the standard deviation of pixel values over the region. The Canny Edge Detection algorithm first smooths the image using a Gaussian filter with $\sigma=1$, then it finds the intensity gradients of the image, next it apply non-maximum suppression to retain the local maximum gradients, finally any pixel with gradient magnitude less than a certain threshold is removed.
    \begin{figure}
        \centering
        \subfloat[SAR VV band]{\includegraphics[width=4cm]{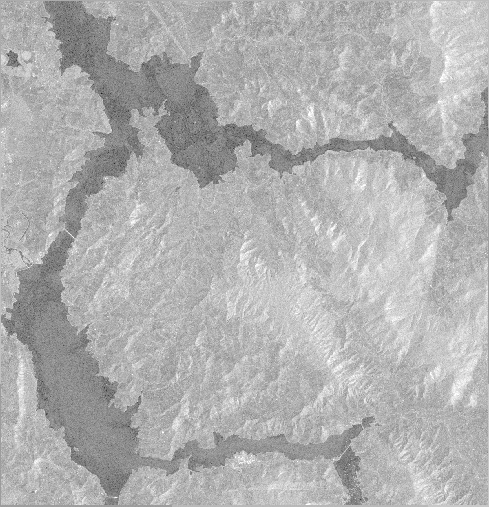}}
        \hfill
        \subfloat[SAR VH band]{\includegraphics[width=4cm]{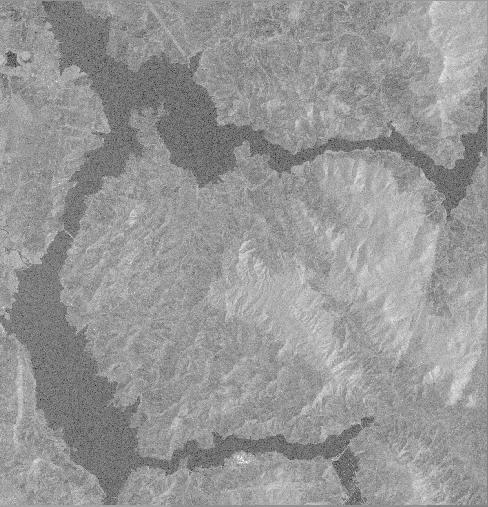}}
        \vspace{0pt}
        \subfloat[Raw combined SAR image]{\includegraphics[width=4cm]{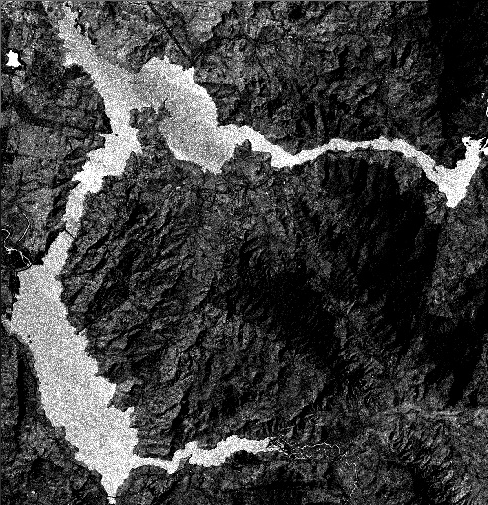}}
        \hfill
        \subfloat[Speckle filtered SAR image]{\includegraphics[width=4cm]{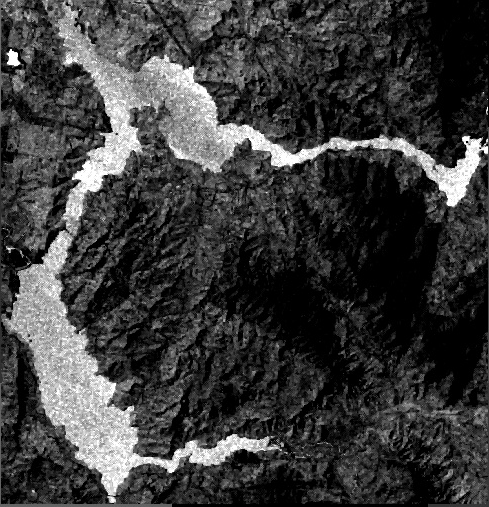}}
        \vspace{0pt}
        \subfloat[Canny Edge Detection result. Blue pixels indicate pixels considered as edges]{\includegraphics[width=8cm]{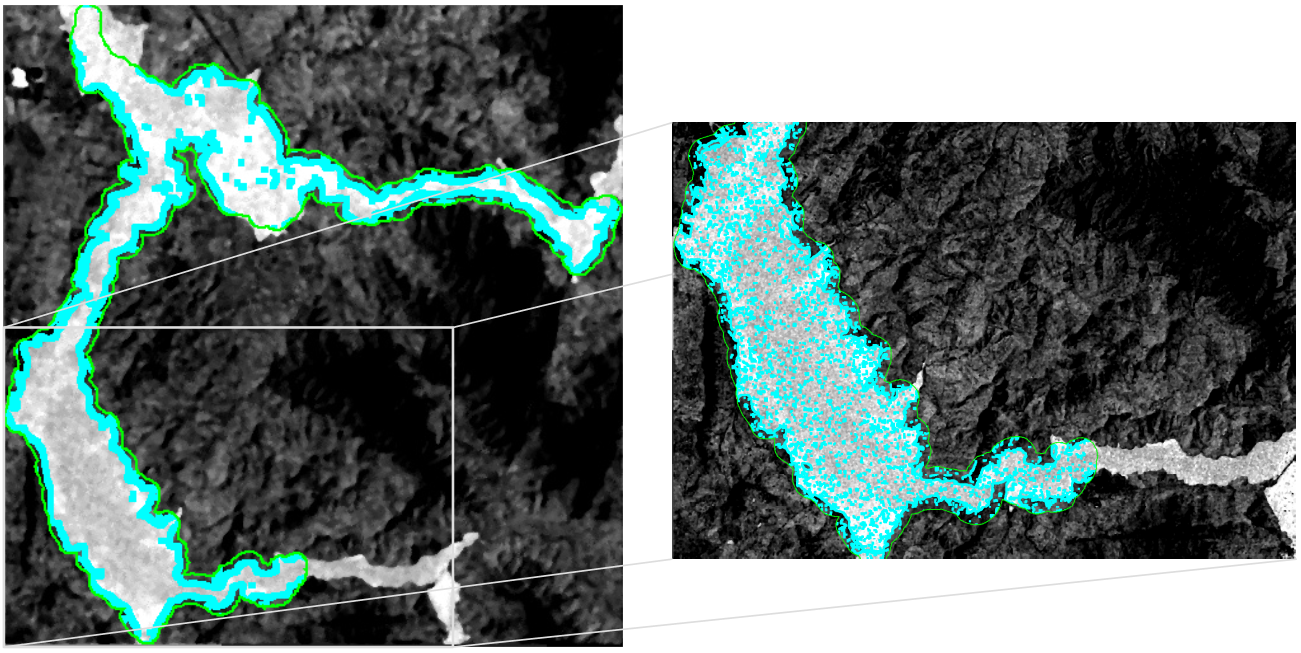}}
        \caption[Preprocessing steps demonstrated for dam Hume on 2018-11-16]{Preprocess steps demonstrated for dam Hume on 2018-11-16.}
        \label{fig:hume_preprocess}
    \end{figure}
\end{enumerate}

For water estimation, we raise the candidate water level from the lowest point to the highest point on the DEM image to see which level matches best with the SAR image. First, we find the minimum and maximum elevation value within the region of the dam, then we evaluate the fitness of each level. However, since the number of levels to be evaluated is not bounded and may be enormous due to it being a continuous variable, we apply a sampling procedure instead.

The fitness function takes in a candidate water level value and returns the "fitness" of that level. More specifically, it simulates the corresponding boundary to that level following algorithm \ref{algo:fitnessfunction}.

\begin{algorithm}
\KwInput{The candidate level}
\KwResult{Fitness value}
DEM\_masked = Mask($DEM \leq level$);\\
DEM\_cc = ExtractConnectedComponents(DEM\_masked);\\
DEM\_bound = BiggestConnectedComponent(DEM\_cc);\\
sum\_grads = Sum(canny pixels along DEM\_bound);\\
\Return sum\_grads;
\caption{Fitness function}
\label{algo:fitnessfunction}
\end{algorithm}
% \begin{enumerate}
%     \item Preliminary mask: every DEM pixel whose value less than or equal to the considering level is seen as water pixel ($mask=1)$, while others are seen as land pixels ($mask=0$).
%     \item Extract connected components: on the binary image on the previous step, we find connected components of water.
%     \item Finally, the biggest connected component from the last step is considered as the simulated dam boundary.
% \end{enumerate}

To take our assumption into account, we build the fitness value as the sum of residual gradient magnitude on the Canny Edge image, along the simulated dam boundary.

Since each evaluation of the fitness function requires at least one traversal of both the DEM and SAR image, it is considered very costly, therefore we must reduce the evaluations as much as possible. Here we employ a sampling technique as in algorithm \ref{algo:waterestimate}.
% \begin{enumerate}
%     \item Sample $num$ equally spaced levels from \textit{lower} to \textit{upper}, where \textit{lower} and \textit{upper} depicts the search range.
%     \item Evaluate the fitness values for every levels sampled.
%     \item Choose the level which yields the highest fitness value, denote this value as $best\-level$.
%     \item Recalculate the distance between samples, denote this value as \textit{step}.
%     \item Update the new \textit{lower} as $best\_level-step$, new \textit{upper} as $best\_level+step$.
%     \item Repeat until $step$ is less than a certain $tolerance$ threshold.
% \end{enumerate}

\begin{algorithm}
\KwInput{lower, upper: max \& max DEM}
\KwResult{Estimated water level}
Preprocess();\\
\Do{$step \leq tolerance$}{
candidates = Linspace(lower, upper, sample\_num);\\
values = Map(Fitness, candidates);\\
best\_level = ArgMax(candidates, values);\\
step = (upper – lower) / (sample\_num – 1);\\
lower = best\_level – step;\\
upper = best\_level + step;\\
}
\Return best\_level;
\caption{Water level estimation algorithm}
\label{algo:waterestimate}
\end{algorithm}

Since every step the searching range ($maxDEM - minDEM$) is reduced by a factor of $\frac{num-1}{2}$ until the $step$ ($\frac{maxDEM - minDEM}{num-1}$) goes below the threshold $tolerance$, the algorithm always finishes after about $log_{\frac{num-1}{2}}{\frac{maxDEM - minDEM}{(num-1)tolerance}}$ iterations.

% Figure \ref{fig:evaluation_demo} demonstrates the process of evaluation of 3 level values 180, 190 and 200 meters on Hume dam. For this scene, the gauge measurement is 190.97 meters. It can be observed that the bound at level 180 is significantly smaller than the bright area of the SAR combined band, while the bound at level 200 loosely captures the area, so they both intersect only a few "edge pixels". Only the bound at level 190 fits the most, hence it crosses many edge pixels. Figure \ref{fig:fitness_maximization} shows the process of fitness function maximization.
% \begin{figure}[H]
%     \centering
%     \subfloat[Level = 180]{\includegraphics[width=8cm]{images/hume_l180vis.png}}
%     \vspace{0pt}
%     \subfloat[Level = 190]{\includegraphics[width=8cm]{images/hume_l190vis.png}}
%     \vspace{0pt}
%     \subfloat[Level = 200]{\includegraphics[width=8cm]{images/hume_l200vis.png}}
%     \caption[Evaluations of 3 simulated levels 180, 190 and 200]{Evaluations of 3 simulated levels 180, 190 and 200. a) DEM with simulated water surfaces showed in grey. b) Simulated dam boundaries showed in red, shapefiles in green. c) Zoom-in of b. d) Simulated dam boundaries showed along with edge pixels.}
%     \label{fig:evaluation_demo}
% \end{figure}

\begin{figure}
    \subfloat[Iteration 1. Min = 152, max = 364, step = 26.5, best level = 205]{\centering\includegraphics[width=8cm]{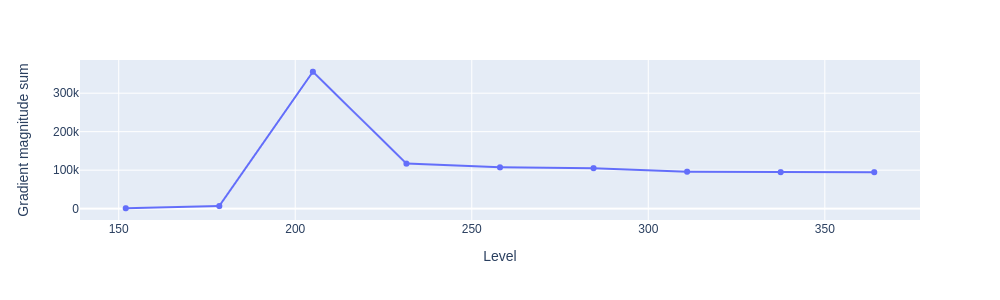}}
    \vspace{0pt}
    \subfloat[Iteration 2. Min = 178.5, max = 231.5, step = 231.5, best level = 191.75]{\centering\includegraphics[width=8cm]{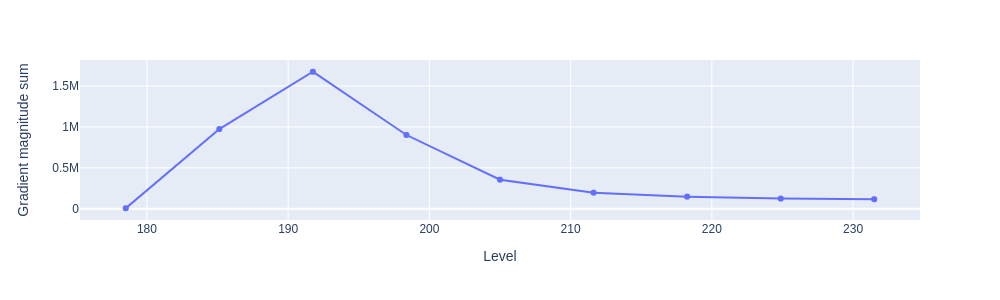}}
    \vspace{0pt}
    \subfloat[Iteration 3. Min = 185.125, max = 198.375, step = 1.65625, best level = 191.75]{\centering\includegraphics[width=8cm]{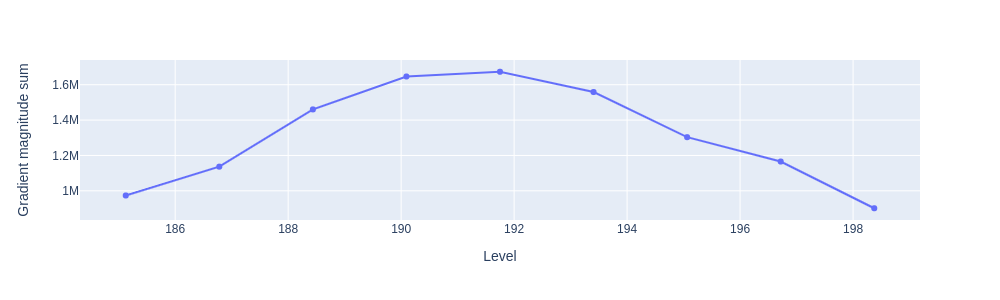}}
    \vspace{0pt}
    \subfloat[Iteration 4. Min = 190.09375, max = 193.40625, step = 0.4140625, best level = 191.3359375]{\centering\includegraphics[width=8cm]{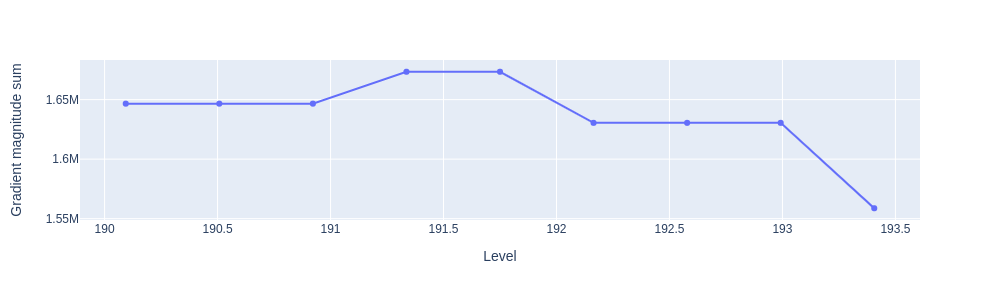}}
    \caption[The process of fitness function maximization of dam Hume on date 2018-11-16]{The process of fitness function maximization of dam Hume on date 2018-11-16. The sample size is 9, tolerance is 1 meter and it takes 4 iterations to find the optimal level.}
    \label{fig:fitness_maximization}
\end{figure}
\section{Experiments}
\label{experiment}
To assess the accuracy of proposed algorithm, we compare the estimated results of 3 dams (Burrendong, Hume and Mosul) versus their reference data. More specifically, Burrendong and Hume reference data are automatic measurements at their gauge stations which can be publicly accessed by WaterNSW \footnote{https://realtimedata.waternsw.com.au}, whereas Mosul data is achieved using Altimetry satellite from Hydroweb \footnote{http://hydroweb.theia-land.fr}. Table \ref{tab:sites_details} shows details about the target sites.

Mentioning the DEM, for site Burrendong we employ the Australian 5M DEM derived from LiDAR model representing a National 5m DEM which has been derived from 236 individual LiDAR surveys between 2001 and 2015 covering an area of over 245,000 $km^2$. Meanwhile, SRTM v3 is used for Hume and Mosul. Further comparisons between those kinds of DEM can be found in table \ref{tab:compare_aus_srtm}.

\begin{table}
    \centering
    \begin{tabular}{|c|c|c|c|}
         \hline
         \textbf{DEM} & \textbf{Resol.} & \textbf{Spatial coverage} & \textbf{Acquisition time} \\ \hline
         AUS & 5m & 245,000 $km^2$ over Australia & 2001-2015 \\ \hline
         SRTM v3 & 30m & Near-global & 2000 \\ \hline
    \end{tabular}
    \caption{Comparison between AUS and SRTM DEMs.}
    \label{tab:compare_aus_srtm}
\end{table}

Since the DEM also captures the water surface at the moment of acquisition, we do not have information about the topography below that surface level, so the algorithm can not infer the water level when it goes lower than that static surface level. The DEM water surface levels for Burrendong, Hume and Mosul are 324, 180 and 307 respectively. Therefore, we only concern about the dates when the reference water levels are higher than the water surface of the corresponding DEM. Among these days, we consider the dates where both satellite observation and reference data are available. Additionally, 8 random dates are selected for calibration while the rest are used for evaluation.

\begin{table}
    \centering
    \begin{tabular}{|c|c|c|c|c|c|}
         \hline
         \textbf{Name} & \textbf{Country} & \textbf{Lat} & \textbf{Lon} & \textbf{Area ($km^2$)} & \textbf{Reference type} \\ \hline
         Burrendong & Australia & -32.67 & 149.11 & 62.2 & Gauge station \\ \hline
         Hume & Australia & -36.11 & 147.03 & 110.94 & Gauge station \\ \hline
         Mosul & Iraq & 36.63 & 42.83 & 346.9 & Altimetry \\ \hline
    \end{tabular}
    \caption{Descriptions of dams used for evaluation.}
    \label{tab:sites_details}
\end{table}

Table \ref{tab:calib_dates} indicates the dates used for calibration for each dam.

\begin{table}
    \centering
    \begin{tabular}{|c|c|c|}
         \hline
         \textbf{Dam} & \textbf{DEM} & \textbf{Calibration dates} \\
         \hline
         \multirow{2}{*}{Burrendong}&\multirow{2}{*}{AUS} & 2018-09-29, 2018-01-20, 2016-10-18, 2016-12-20,\\
         && 2018-05-20, 2017-02-18, 2018-11-16, 2018-07-31 \\ \hline
         \multirow{2}{*}{Hume} &\multirow{2}{*}{SRTM}& 2014-12-15, 2018-11-28, 2015-01-13, 2018-09-17,\\
         && 2016-07-17, 2016-11-14, 2016-07-29, 2017-01-25 \\ \hline
         \multirow{2}{*}{Mosul} & \multirow{2}{*}{SRTM} & 2017-04-02, 2017-06-01, 2015-06-18, 2017-03-21,\\
         && 2019-02-03, 2019-05-23, 2018-06-20, 2017-09-06 \\ \hline
    \end{tabular}
    \caption{8 dates used for calibration for each of 3 dams Burrendong, Hume and Mosul.}
    \label{tab:calib_dates}
\end{table}

\begin{figure}
    \centering
    \subfloat[Burrendong]{\includegraphics[width=9cm]{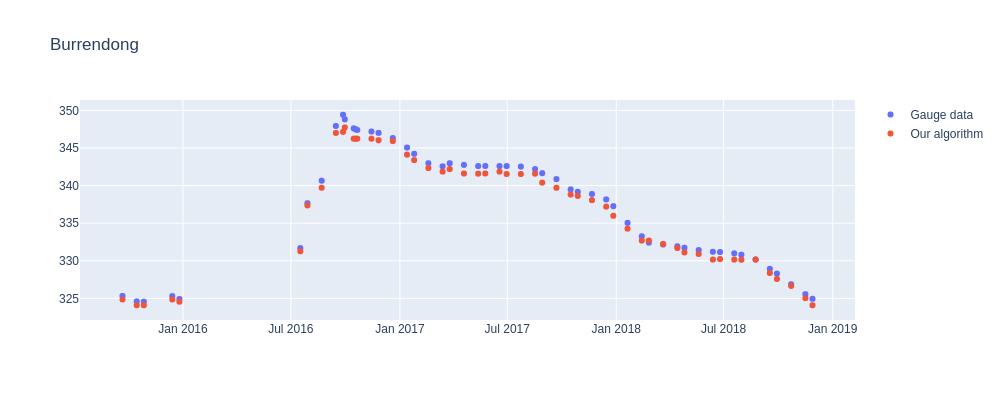}}
    
    \subfloat[Hume]{\includegraphics[width=9cm]{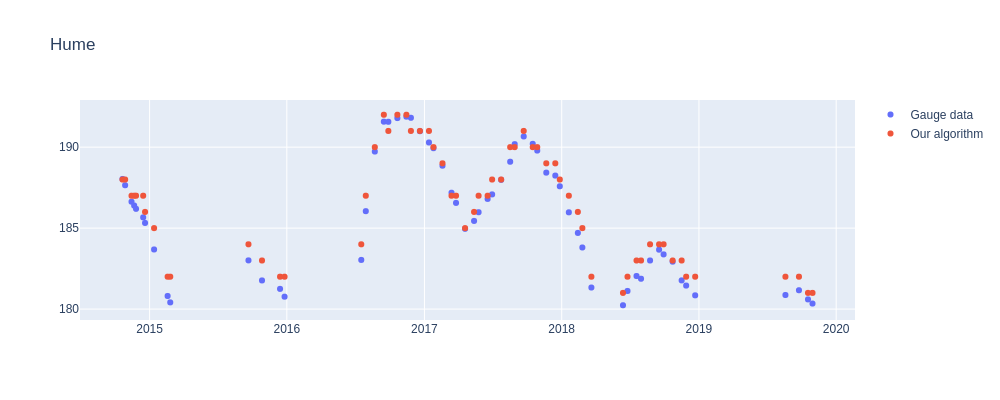}}
    
    \subfloat[Mosul]{\includegraphics[width=9cm]{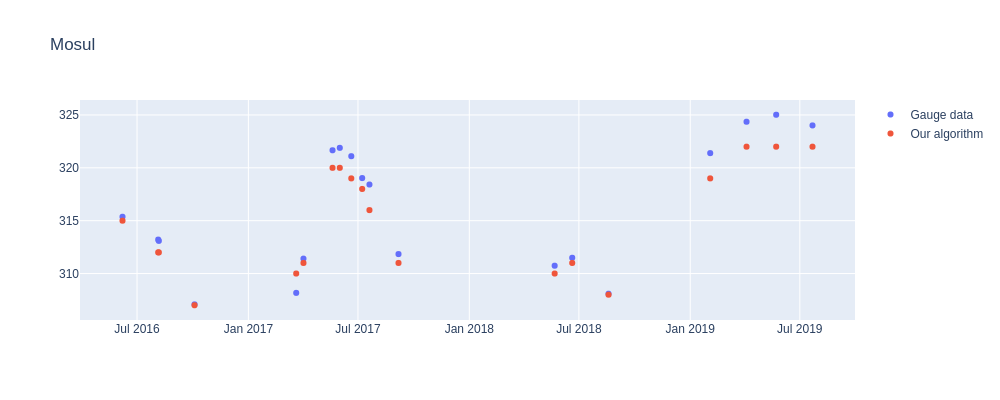}}
    
    \caption{Full time series of proposed algorithm's estimations and reference water levels for all 3 dams.}
    \label{fig:full_timeseries}
\end{figure}

Using the calibrated parameters, we evaluate the correctness of the algorithm over the remaining dates. Table \ref{tab:eval_results} shows the resulted evaluations. We achieved a very high $R^2$ score of over 0.96 among 3 dams. More remarkably, the Mean Absolute Error is less than 1 meter, which is far less than the SRTM error of 3.7m. Figure \ref{fig:full_timeseries} illustrates the full time series of estimated water level versus reference data. To our knowledge from other works on remote sensing, this level of error is acceptable and promising for the use of water level monitoring and thus should be furthermore developed.

\begin{table}
    \centering
    \begin{tabular}{|l|c|c|c|c|}
         \hline
         \textbf{Dam} & \textbf{$R^2$} & \textbf{RMSE} (meters) & \textbf{MAE} (meters) & \textbf{\# of dates} \\ \hline
         Burrendong & 0.99 & 0.88 & 0.78 & 45 \\ \hline
         Hume & 0.97 & 0.77 & 0.66 & 53 \\ \hline
         Mosul & 0.93 & 1.50 & 1.26 & 14 \\ \hline
         \textbf{Mean} & \textbf{0.96} & \textbf{1.09} & \textbf{0.93} & \\ \hline
    \end{tabular}
    \caption{Evaluation results of each dam over the remaining dates.}
    \label{tab:eval_results}
\end{table}

\section{Conclusion and Future Work}
\label{conclusions}
In this work we have analyzed existing works on water level estimating methods and then found a novel approach that no one has taken before. The method involves extracting water body edges and searching for the most likely water elevation on DEM using sampling technique. Next we demonstrated the potential of our method on real-life reference data, with high $R^2$ score of 0.96 on average and low average error of 0.93 meters. To the best of our knowledge, this is the lowest error rate achieved using Sentinel-1 imagery and SRTM data for water level estimating.

With the potential results, we suggest some improvements to our works: to reduce the number of simulated level evaluations, we would like to experiment on the use of other optimization methods such as Bayesian Optimization. We also believe it is beneficial to consider deep learning segmentation networks like U-net for accurately segmenting water bodies.

If all of the above suggestions were effective, we will be able to design a reliable and valuable system that can monitor and predict extreme events (flood \& drought), facilitating appropriate reaction time for farmers and governments. Besides, integrated multiple remote sensing datasets \cite{Cretaux2011SOLS:Data} should be considered. 
\bibliographystyle{IEEEtran}
\nocite{*}
\bibliography{references}

\end{document}